\documentclass[11pt]{article}
\usepackage{graphicx}
\usepackage{amssymb}
\usepackage{epstopdf}
\def\calm{\mathcal{M}}
\def\calt{\mathcal{T}}
\def\call{\mathcal{L}}
\def\cala{\mathcal{A}}
\def\calv{\mathcal{V}}
\def\ab{ab}
\def\prefac{{\Omega_{D-3}\over 16\pi G}}
\def\lprefac{{\Omega_{D-3}\call\over 16\pi G}}
\def\bd{\bar D}
\def\ty{\tilde y}
\def\tz{\tilde z}

\begin{document}

 \begin{titlepage}
 \vspace*{0.6in}
\begin{center}
{\Large\bf Stresses and Strains in the First Law}

\vspace{0.3cm}
{\Large\bf  for Kaluza-Klein Black Holes}

\vspace{1cm}

{\large David Kastor and Jennie Traschen\\
Department of Physics\\
University of Massachusetts\\
Amherst, MA 01003\\
}
\end{center}
\vskip 1cm
\begin{center}\today\end{center}

\vskip 0.6in

\begin{abstract}
{\large We consider how variations in the moduli of the compactification manifold contribute `$pdV$' type work terms to the first law for Kaluza-Klein black holes.  We give a new proof for the $S^1$ case, based on Hamiltonian methods, which demonstrates that the result holds for arbitrary perturbations around a static black hole background.   We further apply these methods to derive the first law for  black holes in $2$-torus compactifications, where there are three real moduli.  We find that the result can be simply stated in terms of constructs familiar from the physics of elastic materials, the stress and strain tensors.  The strain tensor encodes the change in size and shape of the $2$-torus as the moduli are varied.  The role of the stress tensor is played by a tension tensor, which generalizes the spacetime tension that enters the first law in the $S^1$ case.}
   \end{abstract}
    
\vskip 0.6in
\begin{center}
{\Large \textsf{Dedicated to Rafael Sorkin in honor of his 60th birthday}}
\end{center}

\vfill

 \end{titlepage}

\section{Introduction}

In recent years there has been great interest in higher dimensional black hole spacetimes.  The discovery of stationary black rings in five dimensions with horizon topology $S^2\times S^1$ \cite{Emparan:2001wn}, in particular, showed that the classification of asymptotically flat black holes will not be a simple extension of the four dimensional results.  

Interest has also focused on black holes with Kaluza-Klein boundary conditions, {\it i.e.} those that approach Minkowski space $M$ times a compact Ricci-flat space $K$ at infinity.
Already in the simplest case of compactification on a circle, Kaluza-Klein black holes display many interesting physical phenomena, such as the Gregory-Laflamme instability of black strings and the black hole-black string transition\footnote{We refer the reader to the reviews \cite{Harmark:2005pp} and \cite{Kol:2004ww} for detailed discussions and references.}.  For compactification on a more general Ricci-flat compact space $K$, one expects similar phenomena.  A black hole horizon, for example,  may  be localized on 
$K$, or it may wrap some topologically non-trivial cycle of $K$.   The space of black hole solutions with these boundary conditions will exhibit a correspondingly  rich phase structure.

In this paper we will focus on the first law of black hole mechanics for Kaluza-Klein black holes.  For asymptotically flat black holes, the first law \cite{Bardeen:1973gs} relates the variation of the horizon area to the variations of quantities defined at infinity, such as the ADM mass and angular momentum.  For Kaluza-Klein black holes, there are additional parameters that characterize the  spacetime at infinity, namely the moduli of the compact manifold $K$.  One can now consider two nearby black hole solutions that both approach $M\times K$ at infinity, but with slightly shifted values of the moduli.  Our particular object in this paper will be determining how such variations in the moduli enter into the first law for Kaluza-Klein black holes.

If the compact manifold is a circle, there is a single modulus, which is simply the length $\call$ of the circle.  In this case, the first law including variations in $\call$ has been worked out in 
\cite{Townsend:2001rg}\cite{Harmark:2003eg}.  In addition to the length $\call$, spacetimes asymptotic to $M\times S^1$ are 
characterized at  infinity by the ADM mass $\mathcal{M}$ and 
the tension $\mathcal{T}$.  The ADM mass  is, of course, a familiar quantitiy and is defined with respect to the asymptotic time translation Killing vector.  The spacetime tension $\calt$  \cite{Traschen:2001pb}\cite{Townsend:2001rg}\cite{Harmark:2004ch}  is similarly defined with respect to the asymptotic spatial translation Killing vector around the $S^1$.  The first law  for black hole spacetimes  is then given by \cite{Townsend:2001rg,Harmark:2003eg}
\begin{equation}\label{firstlaw}
\delta\calm={\kappa\over 8\pi G}\delta\cala +\calt\delta\call .
\end{equation}
We see that in addition to the usual $\kappa\delta\cala$ term, there is a `work' term given by the product of the tension and the variation in the length of the $S^1$ at spatial infinity.    
A positive spacetime tension corresponds to a negative pressure.  
It was shown in reference \cite{Traschen:2003jm} that the gravitational contribution to the spacetime tension, exclusive of any matter sources which might have negative tension ({\it i.e.} positive pressure),  is always positive.  Hence the coefficient of $\delta\call$ has the opposite sign than is common in thermodynamics, where one generally considers positive pressures.  

In this paper, we will present a  new derivation of the first law (\ref{firstlaw}) for $S^1$ Kaluza-Klein black holes.  
 Our proof  is based on Hamiltonian methods that were used in reference \cite{Traschen:1984bp} to derive a Gauss's law type relation for perturbations in general relativity, and in  reference \cite{Sudarsky:1992ty} to derive the first law with asymptotically flat boundary conditions.
This new proof extends the range of validity of the first law  (\ref{firstlaw}) in the following way.  While the derivations given in \cite{Townsend:2001rg} and \cite{Harmark:2003eg} each make certain symmetry assumptions about the allowed perturbations, the new derivation, like that of  
\cite{Sudarsky:1992ty}, holds for arbitrary perturbations between solutions.   In particular,  although equation  (\ref{firstlaw}) refers to perturbations around a static black hole, the perturbations themselves need not be static.  With our analysis is is also straightforward to add perturbative sources of stress-energy, see section (\ref{sources}).

We then go on to apply the Hamiltonian formalism to derive the first law in the case that the compact space $K$ is a $2$-torus.  We consider how the ADM mass of a black hole varies under arbitrary deformations of the shape and size of the $T^2$.  We find that the first law, in this case, takes the form
\begin{equation}\label{toruslaw}
\delta\calm={\kappa\over 8\pi G}\delta\cala +\calv\, \calt^{AB}\delta\sigma_{AB}.
\end{equation}
Here $\calv$ is the volume of the torus, $\delta\sigma_{AB}$ is the strain tensor corresponding to the deformation of the $T^2$, and $\calt^{AB}$ is a ``tension tensor" that generalizes the tension $\calt$ of the $S^1$ case.  We expect that this result will extend to compactifications on higher dimensional tori $K=T^n$ as well.

\section{ADM mass and tension}
 
We begin by reviewing  the formulas for the ADM mass and tension in the case $K=S^1$ \cite{Harmark:2003eg}.  Let us write the spacetime metric near infinity as $g_{\ab}=\eta_{\ab}+h_{\ab}$, where $\eta_{\ab}$ is the $D$-dimensional  Minkowski metric.  The components of $h_{\ab}$ are assumed to fall-off sufficiently rapidly that the integral expressions for the mass and tension  are well-defined (see equation (\ref{asymptotics}) below).
In the asymptotic region, write the spacetime coordinates as $x^a=(t,z,x^i)$, where $i=1,\dots,D-2$.  The coordinate $z$ goes around the $S^1$ and is identified with period $\call$.  
The ADM mass is the gravitational charge associated with the asymptotic time translation symmetry $\partial/\partial t$.  
If $\Sigma$ is a spatial slice and  $\partial\Sigma_\infty$  its boundary at spatial infinity, then in asymptotically Cartesian coordinates,   the ADM mass is  given by the integral
 \begin{equation}\label{admdefn}
\calm={1\over 16\pi G}\int_{\partial\Sigma_\infty} dz\ ds_i \left( -\partial^ih_j{}^j-\partial^i h_z{}^z +\partial_j h^{ij}\right)
\end{equation}
where indices are raised and lowered with the asymptotic metric $\eta_{\ab}$ and the area element $ds_i$ is that of a sphere $S^{D-3}$ at infinity in a slice of constant $t$ and $z$.  The tension is 
the gravitational charge associated with the asymptotic spatial translation Killing vector $\partial/\partial z$. The tension is similarly given by the integral
\begin{equation}\label{tensiondef}
\calt= - {1\over 16\pi G}\int_{\partial\Sigma_\infty} \ ds_i \left( -\partial^ih_j{}^j-\partial^i h_t{}^t +\partial_j h^{ij}\right).
\end{equation}

Note that in contrast with the ADM mass,  the definition of the tension does not include an integral in the $z$-direction.     
The ADM mass is an integral over the boundary of a slice of constant $t$, which includes the direction around the $S^1$.   The tension, on the other hand,  is defined \cite{Traschen:2001pb}\cite{Townsend:2001rg}\cite{Harmark:2004ch} by an integral over the boundary of a slice of constant $z$.  This includes, in principle,  an integration over time.    However, if one expands the integrand around spatial infinity, one finds that terms that make non-zero contributions to the integral are always time independent.  Time dependent terms fall-off too rapidly to contribute.  Hence, one can omit the integration over the time direction and work with the quantity $\calt$ defined above, which is strictly speaking a  `tension per unit time'.   Similar issues will arise in the definition of the tension tensor, when we consider $T^2$ compactifications.  It is also useful to note, that the expression (\ref{admdefn}) for the ADM mass does not involve $h_{tt}$, while the expression (\ref{tensiondef}) for the tension does not involve $h_{zz}$.  These features follow naturally from the Hamiltonian derivation of the gravitational charges (see {\it e.g.} references \cite{Regge:1974zd}\cite{Kastor:1991ir}\cite{Traschen:2001pb}\cite{Sudarsky:1992ty}).

Making use of the linearized Einstein equations in the asymptotic region,  the ADM mass and tension intergrals may be evaluated  \cite{Harmark:2003eg} in terms of the asymptotic behavior of the metric coefficients
\begin{equation}\label{asymptotics}
g_{tt}\simeq -1 + {c_t\over r^{D-4}},\qquad  g_{zz}\simeq 1 + {c_z\over r^{D-4}}.
\end{equation}
with the results
\begin{equation}\label{massandtension}
\calm = \lprefac ((D-3)c_t -c_z),\qquad \calt=- \prefac((D-3) c_z -c_t).
\end{equation}
We will make use of these formulas below and derive similar expressions for the mass $\calm$ and tension tensor $\calt^{AB}$ for $T^2$ compactifications.

 \section{The first law: two examples}\label{examples}
 
Before proceeding with our derivation of equation (\ref{firstlaw}), we examine in some greater detail how the first law operates in two simple examples, the uniform black string and the 
static Kaluza-Klein bubble.   These examples help build intuition into the physical significance of the work term in the first law.

\subsection{The uniform black string}

The metric for the uniform black string is given by
\begin{equation}
ds^2 = - (1- {c \over r^{D-4}}) dt^2  +dz^2 +(1- {c \over r^{D-4}})^{-1}dr^2+ r^2 d\Omega_{D-3}^2.
\end{equation}
The coordinate $z$ in the $S^1$ direction may be identified with any period $\call$.
In terms of the asymptotic behavior of the metric coefficients in (\ref{asymptotics}) this means that we have $c_t=c$, while $c_z=0$.   
Using (\ref{massandtension}), one finds that the  ADM mass and tension of the black string are given by
\begin{equation}\label{stringmass}
\calm = \lprefac (D-3)\,c,\qquad \calt= \prefac\, c.
\end{equation}
The tension $\calt$  is simply equal to a constant factor, $1/(D-3)$, times the mass per unit length 
$\calm/\call$.

Given that the metric is flat in the $z$-direction and that the period $\call$  may be set freely by hand, it  seems surprising that changes $\delta\call$  in the period around the $S^1$ are part of a non-trivial thermodynamic relation.  In particular, since the mass $\calm$ in (\ref{stringmass}) is linear in $\call$, one might expect that the coefficient of $\delta\call$ in equation (\ref{firstlaw}) should simply be $\calm/\call$.  However, this intuition misses the fact that, if $\call$ is varied with the parameter $c$ being held fixed, the horizon area will also change.  In order to hold $\cala$ fixed, the parameter $ c$ must be varied in a precise way, as $\call$ is varied.  One way to see how the  first law then works out is to express the mass in terms of the horizon area $\cala$ and the period $\call$ as independent variables.

The  black string horizon has spatial topology  $S^{D-3}\times S^1$ and radius $r_h= c^{1/(D-4)}$.  The horizon area is then given by $ \cala= \Omega_{D-3} \call\, r_h^{D-3}$, while the surface gravity is simply
$\kappa = (D-4)/2r_h$.
We can now solve for the parameter $c$ in terms of $\cala$ and $\call$ and substitute into (\ref{stringmass}) to get the expression for the mass
\begin{equation}
\calm = {D-3\over 16\pi G}(\Omega_{D-3} \call)^{1\over D-3}  \cala^{{D-4\over D-3}}
\end{equation}
in which $\cala$ and $\call$ may be regarded as independent variables.  The surface gravity is similarly given in terms of $\cala$ and $\call$ by
\begin{equation}
\kappa = \left({\cala\over \Omega_{D-3}\call}\right)^{{D-4\over D-3}}
\end{equation}
The first law (\ref{firstlaw}) can then verified by the explicit computations
\begin{equation}
\left .{\delta\calm\over\delta\call}\right |_{\delta\cala=0} = {1\over D-3}{\calm\over\call}=\calt,\qquad
\left .{\delta\calm\over\delta\cala}\right |_{\delta\call=0} = 
{D-4\over 16\pi G} (\Omega_{D-3} \call)^{1\over D-3} 
\cala^{-{1\over D-3}} = {\kappa\over 8\pi G}.
\end{equation}
In particular we see that, with the horizon area held fixed, the work required to stretch the length of the $S^1$ by $\delta\call$  is indeed proportional to the tension $\calt$.  

For static, non-uniform black strings, the ADM mass will again be simply proportional to the period $\call$ as in equation (\ref{massandtension}).  In this case, one cannot do the explicit calculations above to verify the first law.  Since we do not know  the solution in the interior, we cannot vary it  in such a way that the horizon area stays fixed as we vary $\call$.  Nevertheless, the first law guarantees that when the horizon area of the black hole is held fixed, the change in $\calm$ under a variation in $\call$ will be  proportional to the tension $\calt$.
This comes comes about, as we will see below, because the first law follows from a relation between a boundary integral at the horizon and one at spatial infinity.  This relation holds for any perturbation of a static solution to a nearby solution of the equation of motions, and does not rely on any detailed knowledge of the unperturbed solution in the interior.

\subsection{The static Kaluza-Klein bubble}

The static Kaluza-Klein bubble is given by the double analytic continuation of the uniform black string
\begin{equation}\label{bubble}
ds^2 =- dt^2 + (1- {c \over r^{D-4}}) dz^2  +{dr^2\over(1- {c \over r^{D-4}}) } + r^2 d\Omega_{D-3}^2.
\end{equation}
The asymptotic behavior of the metric functions in (\ref{asymptotics}) is given by $c_t =0 $ and $c_z=-c$, while the bubble mass and tension are determined according to equation (\ref{massandtension}) to be
\begin{equation}\label{bubblemass}
\calm=\lprefac c,\qquad \calt = (D-3) \prefac c = (D-3) {\calm\over\call}.
\end{equation}

In this case, the $z$ direction is not flat, and it is perhaps less surprising that varying the period $\call$  should enter non-trivially into the first law.  The application of the first law in the KK bubble case, however, is rather different than the uniform black string and  exposes some new features.

The first observation to make is that the KK Bubble has no horizon.  Hence the first law (\ref{firstlaw}) reduces to the statement $\delta\calm = \calt\delta\call$.  This is again in seeming conflict with the fact that the expression for the ADM mass is  linear in the period $\call$, which would lead one to expect that $\delta\calm$ should be proportional to $\calm/\call$, rather than the tension $\calt$.  

However, there is an additional complication that must be taken into account for the KK bubble.  
As we will see below, the derivation of the first law (\ref{firstlaw}) assumes that the only internal boundaries on a spacelike slice are black hole horizons.  If other internal boundaries were present, then the first law would acquire additional terms.  For general values of the parameters
$c$ and $\call$, the KK bubble has a conical singularity at the bubble radius $r_B=c^{1\over D-4}$.
In order for the metric to be smooth at $r_B$,   the period $\call$ of the $z$ coordinate must be fixed  to be
\begin{equation}\label{fixedl}
\call =  {4\pi r_B\over D-4}.
\end{equation}
In order that the first law (\ref{firstlaw}) should apply as stated to the KK bubble, this condition must be preserved by any variations.  Hence, if the period $\call$ is varied, the parameter $c$ must be varied as well.  This can be done by using equation (\ref{fixedl}) to solve for the parameter $c$ in terms of the period $\call$ and substitute into  (\ref{bubblemass})  to get for $\calm$ purely in terms of $\call$,
\begin{equation}\label{masslength}
\calm = \lprefac \left( {(D-4)\call \over 4\pi}\right)^{D-4}
\end{equation}
We can then vary this formula to obtain
\begin{equation}
\delta\calm = (D-3){\calm\over\call}\delta\call = \calt\delta\call
\end{equation}
in agreement with the first law (\ref{firstlaw}).
Had we not imposed the relation (\ref{masslength}), the resulting conical singularity would introduce a new contribution to the first law.  This will be evident from our derivation of the first law in section (\ref{theproof}) below.

\section{Proving the first law using Hamiltonian techniques}\label{theproof}

We now turn to the derivation of the first law (\ref{firstlaw}) using the Hamiltonian methods of references \cite{Traschen:1984bp}\cite{Sudarsky:1992ty}\cite{Kastor:1991ir}\cite{Traschen:2001pb}.   We do this in two steps.  In the first step we consider only variations that preserve
 the length of the $S^1$ at infinity, {\it i.e.} those that have $\delta\call=0$.   In this case, the previous results extend straightforwardly to the present $M^{D-1}\times S^1$ boundary conditions to establish the result (\ref{firstlaw}) with $\delta\call=0$.  We recount the main points in this derivation below, so that we may see what needs to be modified to allow for $\delta\call\neq 0$.  The essential change that must be dealt with in this case is that the variation 
  $\delta g_{zz}$ does not vanish at infinity.  To focus on this new feature, we will take the background spacetime to be static rather than stationary.

\subsection{The first law with $\delta\call=0$}\label{lzero}

Let  $\bar g_{\ab}$ be a static solution to the vacuum Einstein equations that is asymptotic to ${\cal M}^{D-1}\times S^1$ at spatial infinity.   We will call $\bar g_{\ab}$ the unperturbed, or background, metric.  The first law follows from considering how the gravitational Hamiltonian changes under  perturbations to the background metric.

For the Hamiltonian decomposition, we  write the full set of spacetime coordinates coordinates $x^a$ as $x^a=(t,x^\alpha)$, where $x^\alpha=(z,x^i)$, with $i=1,\dots,D-2$.  We write the spacetime metric 
$g_{ab}$ as
\begin{equation}
ds^2=-N^2 dt ^2 +  \gamma_{\alpha\beta}(dx^\alpha+N^\alpha dt)( dx^\beta+N^\beta dt)
\end{equation}
The background metric is static with respect to the Killing vector $\partial/\partial t$.  Hence, in appropriate coordinates the background shift vector vanishes, {\it i.e.} $\bar N^\alpha=0$.
The background lapse and spatial metric are then denoted by $\bar N$ and  $\bar\gamma_{\alpha\beta}$ respectively.  Note that the gravitational momentum of the background, $\bar\pi^{\alpha\beta}$, also vanishes.

Now  consider the metric  $g_{ab}=\bar g_{ab}+\delta g_{ab}$, which is a perturbation of the static background metric.  The perturbed metric $g_{ab}$ is also required to solve the Einstein equations, but is not required to be static.   
We examine how the gravitational Hamiltonian varies under this perturbation.  The Hamiltonian  is given by 
\begin{equation}\label{hamiltonian}
H=\int d^{D-1}x\sqrt{\gamma}\left\{ N\left( - {}^{(D-1)}R +{1\over \gamma}(\pi^{\alpha\beta}\pi_{\alpha\beta}-
{1\over D-2}\pi^2)\right) - 2N_\beta[D_\alpha({1\over\sqrt{\gamma}}\pi^{\alpha\beta})]\right\},
\end{equation}
where $\pi^{\alpha\beta}$ is the gravitational momentum, and ${}^{(D-1)}R$ and $D_\alpha$ are respectively the scalar curvature and covariant derivative operator for the metric 
$\gamma_{\alpha\beta}$ on the spatial slice. 

Let $H\vert_{\bar g_{ab}}$ be the value of the Hamiltonian evaluated on the background metric, and let
$\left .\delta H\right\vert_{\bar g_{AB}}$ be the first variation of the Hamiltonian, again evaluated at the 
background metric.  
Since the gravitational Hamiltonian vanishes on vacuum solutions, and since we are perturbing between vacuum solutions, we  know that both of these quantities vanish.
The equation $\left .\delta H\right\vert_{\bar g_{AB}}=0$, we will see, then relates a boundary integral at infinity to one at the horizon.  This relation yields the first law, as follows.

We now compute the first order variation in the Hamiltonian due to the perturbation to the background metric.  
Since the momentum and shift vector for the background metric vanish, 
the only nonvanishing contributions to $\left .\delta H\right\vert_{\bar g_{AB}}$  come from varying the scalar curvature  of the spatial metric ${}^{(D-1)}R$.  The first  order variation of the Hamiltonian about the reference metric is then given by 
\begin{equation}
\left .\delta H\right\vert_{\bar g_{AB}}= \int d^{D-1}x\sqrt{\bar\gamma}\bar N
\left\{ {}^{(D-1)}\bar R^{\alpha\beta}\delta\gamma_{\alpha\beta}
- \left(\bar\gamma^{\alpha\rho}\bar\gamma^{\beta\sigma}-\bar\gamma^{\alpha\beta}\bar\gamma^{\rho\sigma}\right) \bd_\alpha\bd_\beta\delta\gamma_{\rho\sigma} 
\right\},
\end{equation}
where indices are raised and lowered using the background spatial  metric and $\bar D_\alpha$ is the corresponding covariant derivative operator.  As in the derivation of the Hamiltonian equations of motion, 
integration by parts gives $\left .\delta H\right\vert_{\bar g_{AB}}$ as the sum of two pieces, a  volume term  and a total derivative.  One finds
\begin{equation}\label{split}
\left .\delta H\right\vert_{\bar g_{AB}}= \delta H_1 + \delta H_2
\end{equation}
where the volume term is given by
\begin{equation}\label{volume}
\delta H_1= \int_\Sigma d^{D-1}x\sqrt{\bar\gamma} \left\{ {}^{(D-1)}\bar R^{\alpha\beta}\bar N
-\bd^\alpha\bd^\beta\bar N 
+ \bar\gamma^{\alpha\beta}\bd_\rho\bd^\rho\bar N \right\}\delta\gamma_{\alpha\beta}
\end{equation}
and the total derivative term is given by 
$\delta H_2 = \int d^{D-1}x\sqrt{\gamma}\bd_\alpha B^\alpha$, with 
\begin{equation}
B^\alpha = \left(\bar\gamma^{\alpha\rho}\bar\gamma^{\beta\sigma}-\bar\gamma^{\alpha\beta}\bar\gamma^{\rho\sigma}\right) 
\left( -\bar N\bd_\beta\delta\gamma_{\rho\sigma} + \delta\gamma_{\rho\sigma}\bd_\beta\bar N\right).
\end{equation}

Consider the volume term $\delta H_1$ first.  The steps that we have followed above are those one would follow in deriving the Hamiltonian equations of motion for the background metric.  In particular, the  integrand of (\ref{volume}) is equal to  the Lie derivative of the background  momentum $\bar\pi^{\alpha\beta}$ with respect to the Killing vector $\partial/\partial t$.  Since the reference metric is static, this quantity vanishes and hence $\delta H_1=0$.  The vanishing of the integrand in (\ref{volume}) can also be checked directly using  the relation 
$\bar\nabla_a\bar\nabla_b\xi_c=- \bar R_{bca}{}^d\xi_d$ for  the Killing vector $\xi=\partial/\partial t$.

Using Gauss's law, $\delta H_2$ can be written as a boundary term
\begin{equation}\label{boundary}
\delta H_2=   \int_{\partial\Sigma} d\bar s_\alpha
\left(\bar\gamma^{\alpha\rho}\bar\gamma^{\beta\sigma}-\bar\gamma^{\alpha\beta}\bar\gamma^{\rho\sigma}\right) 
\left( -\bar N\bd_\beta\delta\gamma_{\rho\sigma} + \delta\gamma_{\rho\sigma}\bd_\beta\bar N\right).
\end{equation}
The boundary of the spatial slice $\Sigma$ has two components, an inner boundary at the black hole horizon and an outer boundary at spatial infinity, which we denote by $\partial\Sigma_{BH}$ and $\partial\Sigma_{\infty}$ respectively.  The equation $\delta H\vert_{\bar g_{AB}}=0$, which is satisfied for perturbations between solutions to the equations of motion,  then reduces to the requirement that 
\begin{equation}\label{sumofboundaries}
\delta H_2\vert_{\partial\Sigma_{BH}} +\delta H_2\vert_{\partial\Sigma_{\infty}}=0
\end{equation}
It is shown in reference \cite{Sudarsky:1992ty} that the boundary integral evaluated at the horizon is given by
\begin{equation}\label{innerboundary}
\delta H_2\vert_{\partial\Sigma_{BH}} = 2\kappa\delta\cala
\end{equation}
Although reference \cite{Sudarsky:1992ty} was concerned with asymptotically flat spacetimes, 
this last result just depends on $\partial/\partial t$ being the generator of a Killing horizon\footnote{We also assume, following reference \cite{Sudarsky:1992ty} that the horizon is a bifurcate Killing horizon.}.  
It therefore continues to hold, in particular,  in the case of black holes asymptotic to 
$M^{D-1}\times S^1$.  

Using this result, we can write equation (\ref{sumofboundaries}) as
\begin{equation}\label{almost-there}
{\kappa\delta\cala\over 8\pi G} ={1\over 16\pi G} \int_{\partial\Sigma_\infty} d\bar s^\alpha
\bar\gamma^{\beta\rho}\left\{
\bar N(\bd_\beta\delta\gamma_{\alpha\rho}-\bd_\alpha\delta\gamma_{\beta\rho}) -
(\delta\gamma_{\alpha\beta}\bd_\rho\bar N - \delta\gamma_{\beta\rho}\bd_\alpha\bar N)\right\}
\end{equation}
To evaluate the right hand side of equation (\ref{almost-there}) note that with our boundary conditions, the components of $\delta\gamma_{\alpha\beta}$ all fall-off at least as fast\footnote{When we allow the length $\call$ of the $S^1$ to vary below, the component $\delta\gamma_{zz}$ will also have a constant piece at infinity, which gives the change in length.} as  $1/r^{D-4}$, the fall-off required for finiteness of the ADM mass and tension.  Thus, although the volume element $d\bar s_\alpha$, the derivative operator $\bd_i$ and the lapse $\bar N$ are those of the  background metric $\bar g_{ab}$, only the long distance, flat space limits of these quantities contribute to the integral.  
For example, in evaluating the integral it is accurate to make the replacement 
\begin{equation}
\bar\gamma^{\alpha\beta}\delta\gamma_{\alpha\beta} \simeq \delta^{\alpha\beta}\delta\gamma_{\alpha\beta}.
\end{equation}
Terms involving the derivative of lapse then, in particular,  fall-off too fast to give non-zero contributions.  The right hand side of (\ref{almost-there}) then has the same form as  equation  (\ref{admdefn}) for the ADM mass, with $h_{\alpha\beta}$ replaced by $\delta\gamma_{\alpha\beta}$, and hence gives the change $\delta\calm$ in the ADM mass under the perturbation.
Hence, equation (\ref{almost-there}) becomes the first law
\begin{equation}\label{basiclaw}
\delta\calm={\kappa\delta\cala\over 8\pi G}
\end{equation}
for black holes asymptotic to $M^{D-1}\times S^1$ with the length of the $S^1$ at infinity held fixed.  

We pause to point out how Kaluza-Klein bubbles fit into this framework.  
Equation (\ref{sumofboundaries}) assumes that the only internal boundary is a black hole horizon.  Kaluza-Klein bubbles have no black hole horizons, but a generic bubble will have a conical singularity on the axis of the $\partial/\partial z$ Killing vector.  This means that special care must be used in the Hamiltonian treatment, which effectively introduces an inner boundary surrounding the conical singularity (see reference \cite{Hawking:1995fd}).  Our treatment has implicitly assumed that there are no inner boundaries of this type present.

\subsection{The first law with $\delta\call\neq 0$}\label{whereitsat}

We now allow the length $\call$  of the $S^1$ to vary and examine what changes this introduces to the calculations above.   We begin by recalling the formula for the ADM mass
\begin{equation}\label{massagain}
\calm = \lprefac ((D-3)c_t -c_z),
\end{equation}
which assumes the asymptotic forms of the metric coefficients given in (\ref{asymptotics}).  We can vary the length of the $S^1$ at infinity, preserving these asymptotic forms, by simply changing the period of the coordinate around the $S^1$ to be $\call+\delta\call$.
From equation (\ref{massagain}), it is clear that the change in the ADM mass under this variation is then proportional to $\calm/\call$.  Hence, we can write
\begin{equation}\label{separate}
\delta\calm= \calm {\delta\call\over \call} + \left.\delta\calm\right\vert_{\delta\call=0}
\end{equation}
where, in varying the expression for the mass in (\ref{massagain}), the quantity $\left.\delta\calm\right\vert_{\delta\call=0}$ comes from variations to $c_t$ and $c_z$. 

Let us now return to equation (\ref{sumofboundaries}), which states that the sum of the boundary terms
that contribute to the variation of the Hamiltonian between nearby solutions must vanish.  This statement, which serves as the workhorse for proving the first law, continues to hold with $\delta\call\neq 0$.  Moreover, the boundary term at the black hole horizon continues to be given as in equation (\ref{innerboundary}) by $2\kappa\delta\cala$.  

The boundary term at infinity, however, does change.  Since we are working to linear order in perturbation theory, we can write it as the sum of two pieces.  On the one hand, we have a piece that is not proportional to $\delta\call$.  We have shown above that this piece is given by $16\pi G$ times the variation $ \left.\delta\calm\right\vert_{\delta\call=0}$.   On the other hand, we have a second piece of the boundary term at infinity that is the part proportional to $\delta\call$.  Given this information, equation (\ref{sumofboundaries}) can be written as
\begin{equation}\label{workhorse}
{\kappa\delta\cala\over 8\pi G} = \left.\delta\calm\right\vert_{\delta\call=0} +  \lambda \delta\call .
\end{equation}
The coefficient $\lambda$ is then the quantity that we wish to determine.  We can also go one step further, and substitute for the quantity $\left.\delta\calm\right\vert_{\delta\call=0}$  using equation (\ref{separate}) to get
\begin{equation}\label{nearlythere}
{\kappa\delta\cala\over 8\pi G} = \delta\calm +(\lambda- {\calm\over\call})\delta\call .
\end{equation}

To calculate the contribution $\lambda\delta\call$ to the boundary term at infinity, one must be careful in determining the perturbation $\delta\gamma_{zz}= g_{zz}-\bar g_{zz}$.
The unperturbed background metric has $\bar g_{zz} \simeq 1+ c_z/r^{D-4}$ and the coordinate $z$ around the $S^1$ identified with period $\call$.   In the perturbed metric, the coordinate $\tilde z$ around the $S^1$ is identified with period $\call+\delta\call$ and the corresponding metric component is given by 
\begin{equation}\label{wrongcoord}
g_{\tz\tz}\simeq 1+ {c_z+\delta c_z\over r^{D-4}}.
\end{equation}
In order to compute the perturbation $\delta\gamma_{zz}$, one needs to take the difference of the perturbed and the unperturbed metric expressed in the same coordinate system.
The necessary coordinate transformation is given by 
\begin{equation}
\tz =  z(1+{\delta\call\over\call}),
\end{equation}
so that as in the background the coordinate $z$ has period $\call$.
The behavior of the metric component in the $z$ direction near infinity is given in equation (\ref{asymptotics}).  Transforming the metric to the new coordinate, and dropping terms beyond linear order in the perturbations,  gives 
$g_{zz} = (1+ 2{\delta\call\over\call}) g_{\tz \tz}$, which then lets us determine the metric perturbation to be
\begin{equation}
\delta\gamma_{zz}= 2{\delta\call\over\call}(1+ {c_z\over r^{D-4}})
\end{equation}
Note that this metric perturbation includes both a constant term at infinity, which gives directly the change in length of the $S^1$, and a change in the coefficient of the $1/r^{D-4}$ term.

We now evaluate the contributions proportional to $\delta\call$ in the boundary integral (\ref{boundary}) evaluated at infinity.   In section (\ref{lzero}), we argued that because the metric perturbations fell off like
$1/r^{D-4}$, that the terms involving the derivative of the lapse would fall off too rapidly to contribute, as would terms involving the difference of the derivative operator $\bar D_\alpha$ from the flat derivative operator.  
Since $\delta\gamma_{zz}$ goes to a constant at infinity, this is no longer the case and one finds the following contributions
\begin{eqnarray}
\lambda\delta\call &=& {1\over 16\pi G}\int_{\Sigma_\infty} dz ds_i\left\{
-\partial^i(\bar g^{zz}\delta\gamma_{zz}) + \bar\Gamma_{zz}^i\delta\gamma_{zz}\bar g^{zz}\bar g^{zz}
+\bar g^{zz}\delta\gamma_{zz}\partial^i\bar N\right\}\\
&=& \prefac (D-4)(c_t+c_z) \delta\call, 
\end{eqnarray}
where the behavior near infinity of the lapse function $\bar N$ and the Christoffel symbol  $\bar\Gamma_{zz}^i$ of the reference metric are found from the asymptotic forms (\ref{asymptotics}).
We can now plug our result for $\lambda\delta\call$ into equation (\ref{nearlythere}) and find that the factors combine to give the first law
\begin{equation}
\delta\calm = {\kappa\delta\cala\over 8\pi G} +\calt\delta\call .
\end{equation}
This relation was derived in \cite{Harmark:2003eg} for perturbations that share the symmetries of the background metric.  Here we have shown that the result holds for arbitrary perturbations between solutions.  Further, it is also straightforward within the present formalism to add in perturbative sources of 
stress-energy.

\subsection{Adding source terms to the first law}\label{sources}

Assume as above that $\bar g_{ab}$ is a static vacuum solution.  Also as before, we consider another metric  $g_{ab}=\bar g_{ab} + \delta g_{ab}$ that is perturbatively close to $g_{ab}$.  However, rather than assuming  that the metric $g_{ab}$ solves the vacuum Einstein equations, we now allow for perturbative source terms with stress-energy given by $\delta T_{ab}$.  This alters the analysis above in the following way.  Recall that the Hamiltonian constraint with sources is given by 
$H + 16\pi G\int_\Sigma d^z\, d^{D-2}x\sqrt{\gamma} N \rho=0$, where $\rho$ is the energy density.  Our equation (\ref{sumofboundaries}) then becomes
\begin{equation}
\delta H_2\vert_{\partial\Sigma_{BH}} +\delta H_2\vert_{\partial\Sigma_{\infty}}=-
16\pi G\int_\Sigma dz\, d^{D-2}x\sqrt{\bar\gamma}\bar N \delta\rho
\end{equation}
and hence the first law becomes
\begin{equation}\label{firstlawplus}
\delta\calm={\kappa\over 8\pi}\delta\cala +\calt\delta\call + \int_\Sigma dz\, d^{D-2}x\sqrt{-\bar g}
\delta\rho.
\end{equation}

\section{First law for $T^2$ Kaluza-Klein black holes}\label{for-t2}

Having established the first law for $S^1$ Kaluza-Klein black holes, we now show how the Hamiltonian formalism can be used to derive a first law for black holes in higher dimensional torus compactifications.  We consider explicitly only the $2$-torus.  However, we expect that our results should extend to $n$-torus compactifications as well.

The overall size and shape of a flat $T^2$ are specified by $3$ real moduli.  
We will show that the contribution of variations in these moduli to the first law may be expressed in terms  of the physics of elastic media, {\it i.e.} in terms of stresses and strains.   The work done in making a small deformation of an elastic medium (see {\it e.g.} \cite{landau}) is given by the integral over the body of the contraction of the strain tensor for the deformation with the applied stress tensor.
Varying the moduli of the $T^2$ gives rise to a strain tensor $\delta\sigma_{AB}$, where the indices $A,B$ run over the $2$ coordinates on the torus.  
Further, we will see that the role of the stress tensor is played by the tension tensor $\calt^{AB}$, defined below, which is a natural generalization of the tension $\calt$ of $S^1$ Kaluza-Klein black holes.
Both the tension and strain tensors are constant over the $T^2$.  Hence integration over the torus introduces an overall factor of its volume $\calv$, and
the first law for $T^2$ Kaluza-Klein black holes then takes the form
\begin{equation}
\delta\calm = {\kappa\over 8\pi G} \,\delta\cala+\calv\,  \calt^{AB}\delta\sigma_{AB}.
\end{equation}

One expects that $T^2$ Kaluza-Klein black holes have a rich phase structure 
\cite{Kol:2004pn}\cite{Kol:2006vu}.  A black hole horizon can be localized on the $T^2$,  it may wrap one or the other of the nontrivial cycles, or it may wrap around the entire $T^2$.  There should be critical behavior associated with transitions between all these different types of horizon topologies.   For a fixed shape of the $T^2$ at infinity, in analogy with the work of \cite{Harmark:2003eg} from the $S^1$ case, it is natural to expect that a four-dimensional phase diagram including the mass $\calm$ and the 3 components of $\calt^{ab}$ will be necessary to describe these transitions.  

Moreover, in the $T^2$ case, the details of the phase diagram should additionally depend on the shape 
of the torus.  In the $S^1$ case, there is only a single parameter needed to describe the compactification manifold, namely the length $\call$.  This can be used to set an overall length scale for the axis of the phase diagram as in \cite{Harmark:2003eg}.  Hence, the phase diagram looks the same for any value of $\call$.  For $T^2$ compactifications, the overall volume $\calv$ of the torus may similarly be scaled out.  However, the remaining two real shape parameters for the torus remain.

\subsection{The  tension tensor for $T^2$ Kaluza-Klein black holes}\label{t2-tension}

We define the tension tensor $\calt^{AB}$ for $T^2$ Kaluza-Klein compactifications in the following way.   Consider general relativity linearized around flat space.
The ADM mass $\calm$ is then simply equal to the integral of the mass density $\rho= T_{00}$.  For the $S^1$ compactification of the $z$-direction considered above, the tension $\calt$ in the linearized limit is similarly equal to minus the integral of the pressure $p=T_{zz}$.  In this case the integral is taken over all the spatial directions except the $z$ direction.  We will straightforwardly extend these definitions to obtain a symmetric tensor $\calt^{AB}$ for the $T^2$ case, which we call the tension tensor\footnote{A similar construction has been used in reference \cite{Hovdebo:2006jy} in defining the $(t,z)$ components of the stress-energy tensor of boosted black strings.}.

Assume that the $y$ and $z$ directions are compactified on a $T^2$, which we will specify in more detail below.  Write the full set of spacetime coordinates as $(t,y,z,x^i)$ with $i=1,\dots,D-3$ and assume 
the following asymptotic forms for the metric components, 
\begin{equation}\label{t2asymptotics}
\bar g_{tt}\simeq -1 + {c_t\over r^{D-5}},\qquad  \bar g_{yy}\simeq 1 + {c_y\over r^{D-5}},
\qquad \bar g_{zz}\simeq 1 + {c_z\over r^{D-5}},\qquad \bar g_{yz}\simeq  {b \over r^{D-5}}
\end{equation}
We want to use linearized gravity to obtain formulas analogous to (\ref{massandtension}) for 
$\calm$ and $\calt^{AB}$.

Writing the spacetime metric as $g_{ab}=\eta_{ab}+ h_{ab}$, the linearized Einstein equations are given by
\begin{equation}\label{linear}
2\partial_c\partial_{(a}h_{b)}{}^c - \partial_c\partial^c h_{ab}
-\partial_a\partial_b h_c{}^c = 16\pi G
(T_{ab}-{1\over D-2} \eta_{ab} T_c{}^c)
\end{equation}
where indices are raised and lowered with the flat metric.  Assume that both the sources $T_{AB}$ and the perturbed metric $h_{ab}$ are independent of the $t$, $y$ and $z$ coordinates.  We further assume that the only nonvanishing components of the stress-energy tensor are $T_{00}$, $T_{yy}$, $T_{yz}$ and $T_{zz}$.   By taking various linear combinations of the Einstein equations (\ref{linear}), one can then obtain the following relations
\begin{eqnarray}
{1\over D-5}\partial_k\partial^k \left( (D-4) h_{tt} - h_{yy}-h_{zz}\right) &=& -16\pi G T_{00} \\
{1\over D-5}\partial_k\partial^k \left( (D-4) h_{yy} - h_{tt}+h_{zz}\right) &=& -16\pi G T_{yy} \\ \label{tyy}
{1\over D-5}\partial_k\partial^k \left( (D-4) h_{zz} - h_{tt}+h_{yy}\right) &=& -16\pi G T_{zz} \\ \label{tzz}
\partial_k\partial^k h_{yz} &=& -16\pi G T_{yz} \label{tyz}
\end{eqnarray}

We can now obtain a formula for the ADM mass that is analogous to that in 
equation (\ref{massandtension}), 
\begin{eqnarray}\label{t2mass}
\calm & = & \int dv T_{00} \\
&=& -{\calv\over 16\pi G}  {1\over (D-5)} \int  ds_i \partial^i \left( (D-4) h_{tt} - h_{yy}-h_{zz}\right )\\
& = & {\calv\Omega_{D-4}\over 16\pi G} \left( (D-4)c_t - c_y-c_z)\right). \label{explicitmass}
\end{eqnarray}
We then similarly define the elements of the tension tensor $\calt^{AB}$, with $A,B=y,z$.  
\begin{equation}
\calt^{AB}=- \int d^{D-3}x \, T^{AB},
\end{equation}
where we have omitted the integral over the torus directions as well as the integral over time (see the discussion following equation (\ref{tensiondef})).  This then defines a symmetric tensor.  
Using equatiions (\ref{tyy},\ref{tzz},\ref{tyz}) the volume integrals can be converted to boundary integrals.
Hence the elements of the tension tensor are given explicitly in terms of the coefficients in terms of the 
far field behavior by
\begin{equation}\label{tensiontensor}
 \left(\begin{array}{cc} \calt^{yy} & \calt^{yz} \\
 \calt^{zy} & \calt^{zz}
 \end{array}\right)
=  - {\Omega_{D-4}\over 16\pi G} \left(\begin{array}{cc} 
(D-4)c_y -c_t +c_z & (D-5)b \\
(D-5)b & (D-4)c_z -c_t +c_y
\end{array}\right)
\end{equation}
where the coefficients are defined in (\ref{t2asymptotics}).

\subsection{$T^2$ deformations}\label{t2-deform}

\begin{figure}\label{torusfig}
\centering
\includegraphics[width=4cm]{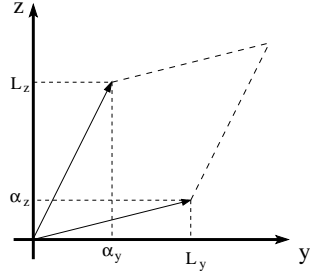}
\caption{A basic cell of the lattice defining the $T^2$ is shown.  The parameters $L_y$, $L_z$, $\alpha_y$ and $\alpha_z$ that give the $y$ and $z$ components of the basis vectors are illustrated.}
\end{figure}

We now turn to the specifics of how we parameterize the shape and size of the $T^2$ at infinity, as well as its deformations.  An $n$-dimensional torus can, of course, be regarded as the Euclidean $n$-plane modded out by a lattice.  For the $2$-torus, the lattice will have $2$ basis vectors and we will use the components of these basis vectors to specify the size and shape of the torus (see figure (\ref{torusfig})).  
The torus is then given by points in the $yz$-plane subject to the identifications
\begin{equation}\label{torusident}
\left(\begin{array}{c} y\\ z\end{array}\right)
\equiv \left(\begin{array}{c} y\\ z\end{array}\right)
+
n_y\left(\begin{array}{c} L_y\\ \alpha_z\end{array}\right) +
n_z\left(\begin{array}{c} \alpha_y \\  L_z\end{array}\right)
\end{equation}
where $n_y,n_z\in Z$.  
For a rectangular torus, in particular, $\alpha_y=\alpha_z=0$ and the parameters $L_y$, $L_z$ are the lengths of the two sides.
Note that we have one more parameter than needed to describe the shape and size of the torus.  The additional degree of freedom describes the orientation of the lattice on the $yz$-plane.  Changing the orientation of the  lattice does not alter the compactification, and hence the extra degree of freedom is unphysical.  Nevertheless, keeping it in the formalism allows us to highlight an interesting point below.

We can now deform the torus slightly by shifting the lattice vectors.  In terms of the components of the lattice vectors, the deformation of the torus is parameterized by
\begin{equation}
(L_y,L_z,\alpha_y,\alpha_z)\longrightarrow (L_y+\delta L_y,L_z +\delta L_z,\alpha_y+\delta\alpha_y,\alpha_z+\delta\alpha_z)
\end{equation}
As we did in the $S^1$ case in section (\ref{whereitsat}), we label the coordinates on the deformed torus with tildes, so that the deformed torus is given by the $\ty\tz$-plane subject to the identifications
\begin{equation}\label{deformed}
\left(\begin{array}{c} \ty\\  \tz\end{array}\right)
\equiv \left(\begin{array}{c} \ty\\  \tz\end{array}\right)
+
n_y\left(\begin{array}{c} L_y+\delta L_y\\ \alpha_z+\delta\alpha_z\end{array}\right) +
n_z\left(\begin{array}{c}\alpha_y+\delta\alpha_y \\ L_z +\delta L_z\end{array}\right).
\end{equation}

We will see that the first law for $T^2$ Kaluza-Klein black holes can be stated succinctly if we treat the torus as an elastic medium (see {\it e.g.} reference \cite{landau}).  In particular, it is useful to compute the strain tensor corresponding to the deformation of the torus described above.   This  is defined in the following way.  If the positions of points in an elastic medium are deformed from their original positions according to the vector field $\xi_A$, then the strain tensor $\delta\sigma_{AB}$  is given by $\delta\sigma_{AB}=\partial_A \xi_B$.

In order to calculate the strain tensor for the deformed $T^2$, we need to be able to identify every point on the deformed torus with a point on the undeformed torus. 
To do this we transform from the coordinates $(\ty,\tz)$ with identifications (\ref{deformed}) back to the original coordinates $(y,z)$ with the identifications (\ref{torusident}).  
The transformation is given by
\begin{equation}\label{tildecoords}
\left(\begin{array}{c} y\\ z\end{array}\right) = \left(\begin{array}{c} \tilde y\\ \tilde z\end{array}\right)
-{1\over \calv} \left(\begin{array}{cc} 
L_z\delta L_y - \alpha_z\delta\alpha_y & -\alpha_y\delta L_y +L_y\delta\alpha_y \\
-\alpha_z\delta L_z +L_z\delta\alpha_z & L_y\delta L_z - \alpha_y\delta\alpha_z
\end{array}\right)\left(\begin{array}{c} \tilde y\\ \tilde z\end{array}\right)
\end{equation}
where $\calv= L_yL_z-\alpha_y\alpha_z$ is the volume of the undeformed torus, and we are working to first order in the deformation parameters $\delta L_y$, $\delta L_y$, $\delta\alpha_y$ and $\delta\alpha_z$. 

We can now think of the deformation of the torus in the following way.  The coordinates  $(y,z)$ specify a  point on the undeformed torus, and  $(\tilde y,\tilde z)$ are the coordinates of the same point on the deformed torus.  The displacement vector $\xi$ is then
given by
\begin{equation}
 \left(\begin{array}{c} \xi_y\\  \xi_z\end{array}\right) =  \left(\begin{array}{c} \ty\\  \tz\end{array}\right) - 
 \left(\begin{array}{c}  y\\  z\end{array}\right), 
\end{equation}
and the strain tensor is found to be
\begin{equation}\label{straintensor}
 \left(\begin{array}{cc}  \delta\sigma_{yy} & \delta\sigma_{yz} \\
 \delta\sigma_{zy} & \delta\sigma_{zz}
 \end{array}\right)
=  {1\over \calv} \left(\begin{array}{cc} 
L_z\delta L_y - \alpha_z\delta\alpha_y & -\alpha_y\delta L_y +L_y\delta\alpha_y \\
-\alpha_z\delta L_z +L_z\delta\alpha_z & L_y\delta L_z - \alpha_y\delta\alpha_z
\end{array}\right)
\end{equation}
Note in particular, that the trace of $\delta\sigma_{AB}$ gives the fractional change in volume of the torus
\begin{equation}
Tr\,\delta\sigma = {\delta \calv\over \calv}.
\end{equation}
The trace-free symmetric part of $\delta\sigma$ gives the shear part of the deformation, while the antisymmetric part gives a rotation of the torus on the $yz$ plane.  As we noted above, such a rotation does not physically alter the torus, and we will see below that indeed the antisymmetric part of the strain tensor does not contribute to the first law.

\subsection{Derivation of the first law}

Let us denote the moduli of the $T^2$ collectively by $\{\rho_I\}=\{L_y,L_z.\alpha_y,\alpha_z\}$ with $I=1,\dots,4$ and we include the unphysical rotational degree of freedom.   If we consider variations around a $T^2$ Kaluza-Klein black hole with the values of the moduli held fixed, $\delta\rho_I=0$, then the results of section (\ref{lzero}) apply and the first law again has the form (\ref{basiclaw}).

We can now consider variations in the moduli as well.  Following the considerations above in  section (\ref{whereitsat}), we can then write an analogue of equation 
(\ref{workhorse}) for the $T^2$ case,
\begin{equation}\label{t2workhorse}
{\kappa\delta\cala\over 8\pi G} = \left.\delta\calm\right\vert_{\delta\rho_l=0} + \sum_{I=1}^4 \lambda_I \delta\rho_I .
\end{equation}
Moreover, the ADM mass depends on the moduli only through the overall volume of the $T^2$, as in equation (\ref{t2mass}).  Therefore, we can write
\begin{equation}
\delta\calm=\left.\delta\calm\right\vert_{\delta\rho_I =0} +  \calm {\delta\calv\over \calv} .
\end{equation}
Combining these last two equations gives
\begin{equation}\label{anotherequation}
{\kappa\delta\cala\over 8\pi G} = \delta\calm + \sum_{I=1}^4 \lambda_I \delta\rho_I - \calm {\delta\calv\over \calv} .
\end{equation}

In order to calculate the contributions to (\ref{t2workhorse}) that come from varying the moduli, we return to equation (\ref{almost-there}), which we reproduce here
\begin{equation}\label{almost-there-again}
{\kappa\delta\cala\over 8\pi G} ={1\over 16\pi G} \int_{\partial\Sigma_\infty} d\bar s^\alpha
\bar\gamma^{\beta\rho}\left\{
\bar N(\bd_\beta\delta\gamma_{\alpha\rho}-\bd_\alpha\delta\gamma_{\beta\rho}) -
(\delta\gamma_{\alpha\beta}\bd_\rho\bar N - \delta\gamma_{\beta\rho}\bd_\alpha\bar N)\right\}.
\end{equation}
We will keep only terms proportional to the variations of the moduli.    We then 
need only consider then, the 
components of the metric perturbation $\delta\gamma_{AB}$ for the torus directions.  
As we did for the $S^1$ case in section (\ref{whereitsat}), 
we must  calculate the perturbation to the metric 
$\delta\gamma_{AB}$ in the original $(y,z)$ coordinate system on the $T^2$, {\it i.e.} in  which the coordinates have the identifications (\ref{torusident}).

Let $(x^1,x^2)=(y,z)$ and $(\tilde x^1,\tilde x^2)=(\tilde y,\tilde z)$.   To linear order in the variations of the moduli, we can then write the coordinate transformation (\ref{tildecoords}) as
\begin{equation}
\tilde x^A=(\delta^A{}_B + \delta\sigma^A{}_B)x^B
\end{equation}
Since we are working to linear order in perturbations and keeping only terms proportional to the variations in the moduli, the metric $\tilde g_{AB}$ on torus in the coordinates $\tilde x^A$ can be taken to equal the background metric $\bar g_{AB}$ in (\ref{t2asymptotics}),
\begin{equation}
\tilde g_{AB}=\bar g_{AB}.
\end{equation}
In this coordinate system, the metric perturbation is encoded in the varied identifications (\ref{deformed}).  Transforming to the coordinates $x^A$ with the original identifications then gives
\begin{equation}
g_{AB}=\tilde g_{AB} +\tilde g_{CB}\delta\sigma^C{}_A +\tilde g_{AC}\delta\sigma^C{}_B.
\end{equation}
The perturbation to the metric $\delta\gamma_{AB}=g_{AB}-\bar g_{AB}$ coming from the deformation of the $T^2$ is then given explicitly by
\begin{eqnarray}
\delta\gamma_{yy}& =& {2\over \calv}\left\{
(1+{c_y\over r^{D-5}})(L_z\delta L_y - \alpha_z\delta\alpha_y) +
{b\over r^{D-5}}(-\alpha_z\delta L_z +L_z\delta\alpha_z)\right\}  \\ \nonumber
\delta\gamma_{yy}& =& {2\over \calv}\left\{
(1+{c_z\over r^{D-5}})(L_y\delta L_z - \alpha_y\delta\alpha_z) +
{b\over r^{D-5}}(-\alpha_y\delta L_y +L_y\delta\alpha_y)\right\}  
\\ \nonumber
\delta\gamma_{yz}=\delta\gamma_{zy} &=& {1\over \calv}  
\left\{ (1+{c_y\over r^{D-5}}) (-\alpha_y\delta L_y +L_y\delta\alpha_y) +
 (1+{c_z\over r^{D-5}}) (-\alpha_z\delta L_z +L_z\delta\alpha_z) \right. \\ && \left. \nonumber
+ {b\over r^{D-5}}(L_z\delta L_y - \alpha_z\delta\alpha_y + L_y\delta L_z - \alpha_y\delta\alpha_z)
\right\}
\end{eqnarray}
In these original coordinates, we see the metric perturbation has both terms that go to constants at infinity and terms that decay like $1/r^{D-5}$.

From the metric perturbation, we can now calculate the right hand side of equation (\ref{t2workhorse}), with the result
\begin{eqnarray}
\sum_{I=1}^4 \lambda_I \delta\rho_I
= {\Omega_{D-4}\over 16\pi G}(D-5)&& \left\{
c_y (L_z\delta L_y-\alpha_z\delta\alpha_y) + c_z(L_y\delta L_z-\alpha_y\delta\alpha_z)
\right . \\ && \nonumber \left.
+ b(L_y\delta\alpha_y-\alpha_y\delta L_y)
+ b(L_z\delta\alpha_z-\alpha_z\delta L_z) + c_t  \delta \calv 
 \right\}
\end{eqnarray}
This result in turn can be plugged into equation (\ref{anotherequation}), which after making use of 
equation (\ref{explicitmass}) can be processed into the form
\begin{eqnarray}\label{almostt2}
\delta M &= &{\kappa\delta\cala\over 8\pi G}  +  {\Omega_{D-4}\over 16\pi G}\left\{ -((D-4)c_y-c_t +c_z)  (L_z\delta L_y-\alpha_z\delta\alpha_y)\right. \\ \nonumber && \left.
 - ((D-4)c_z-c_t +c_y)(L_y\delta L_z-\alpha_y\delta\alpha_z) 
-(D-5)b(L_y\delta\alpha_y-\alpha_y\delta L_y)  \right. \\ \nonumber && \left.
- (D-5)b(L_z\delta\alpha_z-\alpha_z\delta L_z) \right\} .
\end{eqnarray}
This last expression looks quite complicated.
However, if one looks at equation (\ref{almostt2}) with equations (\ref{straintensor}) and (\ref{tensiontensor}) for the strain and tension tensors in mind, one sees that equation (\ref{almostt2}) can be rewritten in the simple form
\begin{equation}\label{done}
\delta M = {\kappa\delta\cala\over 8\pi G} +\calv\, \calt^{AB}\delta\sigma_{AB}.
\end{equation}
This completes the derivation of the first law for $T^2$ Kaluza-Klein black holes\footnote{Based on the first law (\ref{firstlaw}) for $S^1$ Kaluza-Klein black holes, one might be surprised by the prefactor of $\calv$that appears in the work term for the $T^2$ case.  However, if one computes the strain of the $S^1$ following the definitions of section (\ref{t2-deform}), one finds that $\delta\sigma=\delta\call/\call$.  Hence, the work term in the $S^1$ case, when expressed in terms of the strain, becomes $\call\, \calt\delta\sigma$ in parallel with the $T^2$ case.}.  
Recall that the antisymmetric part of the strain tensor represents a rotation of the basic cell of the $T^2$ within the Euclidean plane, leaving its size and shape unchanged.    This mode is present because we kept all four of the components of the lattice basis vectors in our formalism, which is one more degree of freedom that needed.  
It follows from the first law, however, that because the tension tensor is symmetric, the antisymmetric part of the strain tensor makes no contribution.  

\section{Conclusions}

In this paper we have studied how variations in the moduli enter the first law for Kaluza-Klein black holes.
Following the introduction and some basic formulas in the opening sections, in section (\ref{examples}) 
we discussed two simple examples, the uniform black string and the Kaluza-Klein bubble, that illustrate how the first law for Kaluza-Klein black holes operates in practice.   The formula for the ADM mass in these spacetimes is simply proportional to the length of the $S^1$.  However, when this length is varied we saw that holding the horizon area fixed for the black string, or preventing a conical singularity for the bubble, does indeed yield a change in mass proportional to the tension, in accordance with the first law.

In section (\ref{theproof}) we presented a derivation of the first law (\ref{firstlaw}) for $S^1$ Kaluza-Klein black holes  \cite{Townsend:2001rg,Harmark:2003eg} based on the Hamiltonian methods of references \cite{Traschen:1984bp,Sudarsky:1992ty}.  We  highlighted how the $\calt\delta\call$ work term arises from a careful treatment of the boundary term at infinity in equation (\ref{almost-there}).  This careful treatment was necessary because when the moduli are varied, certain components of the metric perturbations have constant pieces at infinity.

In section (\ref{for-t2}) we used the Hamiltonian methods to derive a new result, the first law (\ref{toruslaw}) for $T^2$ Kaluza-Klein black holes.  We saw that the statement of the first law, in this case, takes a simple form if we consider the $T^2$ to be an elastic body, which is deformed by varying the moduli.  The resulting work term is given by the contraction of the strain tensor for the deformed $T^2$ with the tension tensor, which was defined in section (\ref{t2-tension}), times a factor of the overall volume of the $T^2$.

Two interesting directions to pursue further are the following  First, if we consider  $T^n$ Kaluza-Klein black holes with $n>2$, it seems quite likely 
that the first law will in general take the form (\ref{toruslaw}) that we have found in the $T^2$ case.  However, the derivation we have given in the $T^2$ case, involving the explicit representation of the moduli given in section (\ref{t2-deform}), will likely prove cumbersome to extend to the general case.
It would be nice to find a more streamlined derivation.  

Second, we would like to consider more general Ricci-flat compactifications, such as Calabi-Yau manifolds.  In this case, we would also expect the results to be phrased in terms of a strain tensor that encodes the deformation of the manifold as the moduli are varied.  However, the definition of the strain tensor used in section (\ref{t2-deform}) makes use of the representation of a $T^2$ as the Euclidean plane identified under a lattice of translations.
To handle non-flat spaces, a more intrinsic definition of the strain tensor in terms of metric perturbations is necessary.  We plan to return to these topics in future work.

\subsection*{Acknowledgments}

The authors would like to thank Henriette Elvang and Niels Obers for helpful conversations on this topic.  
This work was supported in part by NSF grants PHY-0244801 and PHY-0555304.


\begin{thebibliography}{99}

\bibitem{Emparan:2001wn}
  R.~Emparan and H.~S.~Reall,
  ``A rotating black ring in five dimensions,''
  Phys.\ Rev.\ Lett.\  {\bf 88}, 101101 (2002)
  [arXiv:hep-th/0110260].
  


\bibitem{Harmark:2005pp}
  T.~Harmark and N.~A.~Obers,
  ``Phases of Kaluza-Klein black holes: A brief review,''
  arXiv:hep-th/0503020.

\bibitem{Kol:2004ww}
  B.~Kol,
  ``The phase transition between caged black holes and black strings: A
  review,''
  Phys.\ Rept.\  {\bf 422}, 119 (2006)
  [arXiv:hep-th/0411240].
  
\bibitem{Bardeen:1973gs}
  J.~M.~Bardeen, B.~Carter and S.~W.~Hawking,
  ``The four laws of black hole mechanics,''
  Commun.\ Math.\ Phys.\  {\bf 31}, 161 (1973).
  

  
\bibitem{Townsend:2001rg}
  P.~K.~Townsend and M.~Zamaklar,
  ``The first law of black brane mechanics,''
  Class.\ Quant.\ Grav.\  {\bf 18}, 5269 (2001)
  [arXiv:hep-th/0107228].

\bibitem{Harmark:2003eg}
  T.~Harmark and N.~A.~Obers,
  ``Phase structure of black holes and strings on cylinders,''
  Nucl.\ Phys.\ B {\bf 684}, 183 (2004)
  [arXiv:hep-th/0309230].


\bibitem{Traschen:2001pb}
  J.~H.~Traschen and D.~Fox,
  ``Tension perturbations of black brane spacetimes,''
  Class.\ Quant.\ Grav.\  {\bf 21}, 289 (2004)
  [arXiv:gr-qc/0103106].


\bibitem{Harmark:2004ch}
  T.~Harmark and N.~A.~Obers,
  ``General definition of gravitational tension,''
  JHEP {\bf 0405}, 043 (2004)
  [arXiv:hep-th/0403103].
  
\bibitem{Traschen:2003jm}
  J.~H.~Traschen,
  ``A positivity theorem for gravitational tension in brane spacetimes,''
  Class.\ Quant.\ Grav.\  {\bf 21}, 1343 (2004)
  [arXiv:hep-th/0308173].

\bibitem{Traschen:1984bp}
  J.~H.~Traschen,
  ``Constraints On Stress Energy Perturbations In General Relativity,''
  Phys.\ Rev.\ D {\bf 31}, 283 (1985).


\bibitem{Sudarsky:1992ty}
  D.~Sudarsky and R.~M.~Wald,
  ``Extrema of mass, stationarity, and staticity, and solutions to the Einstein
  Yang-Mills equations,''
  Phys.\ Rev.\ D {\bf 46}, 1453 (1992).

\bibitem{Regge:1974zd}
  T.~Regge and C.~Teitelboim,
  ``Role Of Surface Integrals In The Hamiltonian Formulation Of General
  Relativity,''
  Annals Phys.\  {\bf 88}, 286 (1974).



\bibitem{Kastor:1991ir}
  D.~Kastor and J.~H.~Traschen,
  ``Linear instability of nonvacuum space-times,''
  Phys.\ Rev.\ D {\bf 47}, 480 (1993).
  
\bibitem{Hawking:1995fd}
  S.~W.~Hawking and G.~T.~Horowitz,
   ``The Gravitational Hamiltonian, action, entropy and surface terms,''
  Class.\ Quant.\ Grav.\  {\bf 13}, 1487 (1996)
  [arXiv:gr-qc/9501014].

  
\bibitem{landau}
L.D. Landau and E.M. Lifshitz, "Theory of elasticity," Pergamon Press (1970).


\bibitem{Kol:2004pn}
  B.~Kol and E.~Sorkin,
  ``On black-brane instability in an arbitrary dimension,''
  Class.\ Quant.\ Grav.\  {\bf 21}, 4793 (2004)
  [arXiv:gr-qc/0407058].

\bibitem{Kol:2006vu}
  B.~Kol and E.~Sorkin,
  ``LG (Landau-Ginzburg) in GL (Gregory-Laflamme),''
  arXiv:hep-th/0604015.


  
\bibitem{Hovdebo:2006jy}
  J.~L.~Hovdebo and R.~C.~Myers,
  ``Black rings, boosted strings and Gregory-Laflamme,''
  Phys.\ Rev.\ D {\bf 73}, 084013 (2006)
  [arXiv:hep-th/0601079].


\end{thebibliography}
 \end{document}